\documentclass[journal]{IEEEtran}

\usepackage[utf8]{inputenc}

\usepackage[table,xcdraw]{xcolor}

\usepackage{graphicx}

\usepackage{enumitem}

\usepackage{placeins}

\usepackage[labelformat=parens,subrefformat=parens]{subcaption}

\usepackage[binary-units]{siunitx}

\usepackage{amsfonts}

\usepackage[square,numbers,sort&compress]{natbib}

\usepackage{soul}

\usepackage{diagbox}

\usepackage[textsize=tiny]{todonotes}

\usepackage{lipsum}

\usepackage{float}

\usepackage{stfloats} 

\usepackage{algorithm2e}

\usepackage{algorithmic}

\usepackage{amsmath}

\usepackage[hidelinks]{hyperref}

\usepackage[binary-units]{siunitx}

\hypersetup{breaklinks=true}

\usepackage[labelfont=bf]{caption}

\DeclareSIUnit [ inter-unit-product = ] \Wh { \watt \hour }

\begin{document}

\title{Extreme Coverage in \texttt{5G} Narrowband \texttt{IoT}: \\ a \texttt{LUT}-based Strategy to Optimize Shared Channels}

\author{Emmanuel~Luj\'an,
        Juan~A.~Zuloaga~Mellino,
        Alejandro~D.~Otero,
        Leonardo~Rey~Vega,
        Cecilia~G.~Galarza,
        and~Esteban~E.~Mocskos
\thanks{All authors are with Centro de Simulaci\'on Computacional para Aplicaciones Tecnol\'ogicas, CONICET (CSC-CONICET), Argentina. E-mail: emocskos@csc.conicet.gov.ar }
\thanks{E.~Luj\'an~and~E.~E.~Mocskos are with Departamento de Computaci\'on, FCEyN, Universidad de Buenos Aires, Argentina.}
\thanks{A.~D.~Otero,~L.~R.~Vega~and~C.~G.~Galarza are with Facultad de Ingenier\'ia, Universidad de Buenos Aires, Argentina.}
}

\markboth{IEEE Internet of Things Journal}%
{Luj\'an \MakeLowercase{\textit{et al.}}: Extreme coverage in \texttt{5G} Narrowband \texttt{IoT}: a \texttt{LUT}-based strategy to optimize shared channels of IEEEtran.cls for IEEE Journals}

\maketitle

\begin{abstract}

One of the main challenges in \texttt{IoT} is providing communication support to an increasing number of connected devices. 
In recent years, narrowband radio technology has emerged to address this situation: Narrowband Internet of Things (\texttt{NB-IoT}), which is now  part of \texttt{5G}.
Supporting massive connectivity becomes particularly demanding in extreme coverage scenarios such as underground or deep inside buildings sites.
We propose a novel strategy for these situations focused on optimizing \texttt{NB-IoT} shared channels through the selection of link parameters: modulation and coding scheme, as well as the number of repetitions. 
These parameters are established by the base station (\texttt{BS}) for each block transmitted until reaching a target block error rate (\texttt{BLER$_t$}).
A wrong selection of these magnitudes leads to radio resource waste and a decrease in the number of possible concurrent connections.
Specifically, our strategy is based on a look-up table (\texttt{LUT}) scheme which is used for rapidly delivering the optimal link parameters given a target \texttt{QoS}.
To validate our proposal, we compare with alternative strategies using an open source \texttt{NB-IoT} uplink simulator.
The experiments are based on transmitting blocks of $256$ bits using an \texttt{AWGN} channel over the \texttt{NPUSCH}.
Results show that, especially under extreme conditions, only a few options for link parameters are available, favoring robustness against measurement uncertainties. 
Our strategy minimizes resource usage in all scenarios of acknowledged mode and remarkably reduces losses in the unacknowledged mode, presenting also substantial gains in performance.
We expect to influence future \texttt{BS} software design and implementation, favoring connection support under extreme environments.
\end{abstract}

\begin{IEEEkeywords}
\texttt{NB-IoT}, \texttt{5G}, extreme coverage, lookup table, \texttt{LUT}, physical shared channels
\end{IEEEkeywords}

\IEEEpeerreviewmaketitle

\section{Introduction}

\IEEEPARstart{I}{nternet of Things} (\texttt{IoT}) paradigm is the seamless integration of potentially any object with the Internet~\cite{Li2017}.
Currently, there are more than $16$ billion connected devices worldwide and $2.2$ connected devices per person with an expected growth to nearly $28$ billion connections and $3.4$ networked devices per capita in 2020~\cite{Beyene2017}.
This context imposes a strong need for increasing communication networks capacity. 
Low power wireless technologies have the potential to connect $60\%$ of the devices to the Internet~\cite{keysight}, alleviating the need for further costs in standard network infrastructure.
These technologies, known as low power wide area network (\texttt{LPWAN}), are designed according to the following principles: communication distance up to \SI{40}{\kilo\m}, thousands of devices supported by each base station (\texttt{BS}), availability for over a decade without battery replacement, and module price below U\$D 5~\cite{Lee2017}.

Within \texttt{LPWAN} technologies, the 3rd Generation Partnership Project (\texttt{3GPP}) introduced Narrowband Internet of Things (\texttt{NB-IoT}), a novel narrowband radio technology specifically designed for \texttt{IoT}~\cite{Yu2017} and low-cost deployment.
\texttt{NB-IoT} can satisfy the requirements of non-latency-sensitive and low-bitrate applications (time delay of uplink can be extended to more than \SI{10}{\s}, and uplink or downlink for a single user are supported at \SI{160}{\bit\per\s} at least), with coverage enhancement (coverage capacity is increased \SI{20}{\decibel}), ultra-low power consumption (a \SI{5}{\Wh} battery can be used by one terminal for $10$ years), and massive terminal connections (a single sector can support $50000$ links) with transmission bandwidth of \SI{180}{\kilo\hertz}~\cite{Miao2017}. 
This technology can be directly deployed in Long-Term Evolution (\texttt{LTE}) networks in order to reduce deployment costs~\cite{Yu2017}.
Moreover, \texttt{3GPP} recognizes the importance of this technology describing the open issues to be addressed during the future \texttt{5G} standardization process~\cite{TR23724}.
To address these points, continuously growing efforts were made, mainly over the last two years: radio resource optimization of physical channels such as NPRACH~\cite{Lin2016,Harwahyu2018}, energy efficient uplink scheduling schemes~\cite{Liu2014,Lee2017}, among others.

In this work, we focus on one of the key challenges in \texttt{NB-IoT} and \texttt{5G} environment: to extend the coverage to a massive amount of user equipment (\texttt{UE}) deployed in extreme scenarios. 
Most of the related literature is focused on analyzing coverage performance and comparing it with other \texttt{LPWAN} technologies.
Only some authors propose improvements to the coverage-related issues.
Chafii et al.~\cite{Chafii2018} use a dynamic spectrum based on machine learning techniques for coverage enhancement and energy consumption reduction. 
In particular, the random selection procedure was replaced by a more efficient selection method that chooses the channels with the highest probability to be available, with the best coverage and with the lowest number of repetitions. 
Kocak et al.~\cite{Kocak2017} show how \texttt{UEs} with limited battery life, under extended coverage, can reduce power consumption through payload transmissions savings. 
Particularly, by means of narrowband resource allocations, the use of preamble acknowledgments, and low power targets. 
Concomitantly, Yu et al.~\cite{Yu2017} propose an iterative uplink channel adaptation scheme obtaining significant savings in radio resource consumption. 
In line with these growing efforts, we propose a novel strategy focused on channel link adaptation enhancement, extending the work proposed by Yu et al.~\cite{Yu2017} to extreme coverage scenarios.

For each block transmitted from a \texttt{UE}, the \texttt{BS} adapts the communication link in terms of modulation and coding scheme~(\texttt{MCS}), as well as the number of repetitions~(\texttt{NR}), with an associated cost in radio resources.
The link adaptation finishes when reaching a target block error rate (\texttt{BLER$_t$}) or range of \texttt{BLER} values which defines a target \texttt{QoS}. 
Here we present a strategy for increasing the number of connections in \texttt{NB-IoT} channels subject to this link adaptation process, namely the physical uplink shared channel (\texttt{NPUSCH}) and physical downlink shared channel (\texttt{NPDSCH}). 
In order to achieve this goal, the number of the link adaptation iterations is reduced to a minimum, replacing this process by the use of precalculated values, i.e. a lookup table (\texttt{LUT}).
An open source \texttt{NPUSCH} uplink simulator was developed to assess the performance of this strategy compared with alternative techniques.

In summary, we present:
\begin{itemize}
	\item A strategy for increasing connection massiveness in \texttt{NB-IoT} shared channels, based on decreasing resource consumption.
	\item A \texttt{LUT}-scheme for minimizing link adaptation iterations.
	\item An analysis of this strategy under extreme coverage scenarios, comparing it against alternative strategies.
	\item An \texttt{NPUSCH} uplink simulator used to test the proposed strategy.
\end{itemize}

This article is organized as follows: in sections~\ref{sec:uplink-intro} and~\ref{sec:uplink} an introduction of \texttt{NB-IoT} uplink as well as link adaptation is presented. 
In section~\ref{sec:uplink-scheduling}, we present our \texttt{LUT}-based strategy jointly with other alternative strategies. 
Section~\ref{sec:simulation} reports a brief description of the developed simulation tool and details about the performed experiments. 
Then, in section~\ref{sec:results-and-discussion} results are presented and discussed. 
Finally, conclusions are drawn in section~\ref{sec:conclusions}.

\section{\texttt{NB-IoT} Uplink} \label{sec:uplink-intro}

In this section, we introduce the main concepts related to the \texttt{NB-IoT} uplink. 
\texttt{NB-IoT} supports three deployment modes: independent deployment mode, guard-band deployment mode, and in-band deployment mode; which utilizes a physical resource block (\texttt{PRB}) of \texttt{LTE} carrier wave~\cite{Miao2017}.

The physical random access channel (\texttt{NPRACH}) and the physical uplink shared channel (\texttt{NPUSCH}) are the two specialized uplink channels used to exchange information between an \texttt{UE} and a \texttt{BS}. 
These channels are multiplexed in time or frequency over the assigned bandwidth. 
On the other hand, the downlink channels are the physical broadcast channel (\texttt{NPBCH}), the physical downlink control channel (\texttt{NPDCCH}), and the physical downlink shared channel (\texttt{NPDSCH})~\cite{Beyene2017}. 

The uplink transmission bandwidth is \SI{180}{\kilo\hertz} (regardless of the deployment mode) and two sub-carrier spacings are supported, \SI{3.75}{\kilo\hertz} and \SI{15}{\kilo\hertz}. 
In \SI{3.75}{\kilo\hertz}, single sub-carrier is adopted. 
In \SI{15}{\kilo\hertz}, single or multiple sub-carrier transmission can be adopted~\cite{Miao2017}. 
For both single sub-carrier and multiple sub-carrier, the uplink uses single carrier frequency division multiple accesses (\texttt{SC-FDMA}). 
For \SI{15}{\kilo\hertz} sub-carrier spacing, the \texttt{NB-IoT} uplink frame structure (frame size and time slot length) is the same as the \texttt{LTE} network (\SI{0.5}{\ms} slot, and \SI{1}{\ms} subframe). 
In \SI{3.75}{\kilo\hertz} sub-carrier spacing there is a newly defined time slot of \SI{2}{\ms}. 
One radio frame contains five narrowband time slots and each narrowband time slot contains seven symbols~\citep{Miao2017,Yu2017}.
Both, single sub-carrier and multi sub-carrier should be supported by the \texttt{UEs}, nevertheless, some devices may not support multi sub-carrier mode in the first implementation phase of \texttt{NB-IoT} systems. 
On the other hand, single sub-carrier mode should always be supported~\citep{Yu2017}.

\section{\texttt{NB-IoT} Link Adaptation}\label{sec:uplink}

In this section, we shortly describe the uplink data transfer process and highlight the elements that will be optimized in the following sections.
\texttt{NB-IoT} uplink communication starts with a request from a \texttt{UE} to the \texttt{BS} using the \texttt{NPRACH}. 
Once the \texttt{BS} receives the request for transmission, it returns a scheduling grant to the device including the time and frequency resources allocated. 
Subsequently, the \texttt{UE} transmits a sequence of transfer blocks (\texttt{TBs}) to the \texttt{BS}. 
Algorithm~\ref{alg:BS-THREAD} depicts this process from the \texttt{BS} point of view. 
At the beginning of the uplink communication, the size of the buffer which contains the data that will be transmitted is informed to the \texttt{BS}. 
While this buffer is not empty (line 1), the \texttt{BS} determines the next tuple \texttt{(MCS,NR)} based on the previous one and the \texttt{BLER$_t$} (line 2). 
This is sent to the \texttt{UE} (lines 3 and 4) in the downlink control information (\texttt{DCI}) in conjunction with other control data (like the \texttt{NACK} of the last \texttt{TB}). 
When the \texttt{BS} receives the new \texttt{TB} with a piece of the data buffer, transmitted through the \texttt{NPUSCH} (line 6), it checks if the block has arrived without errors (line 7). 
The simulator developed for this work implements this link adaptation process. 
The function \texttt{nextMCS\&NR()} encapsulates the implementation of our proposed \texttt{LUT} strategy, as well as the other evaluated strategies.
Notice that the function \texttt{nextMCS\&NR()} returns the tuple \texttt{(MCS,NR)}.

\begin{algorithm}
{\fontsize{8.5}{8.5}\selectfont
\begin{algorithmic}[1]
\WHILE{not bufferTransmitted()}
\STATE mcs,nr = nextMCS\&NR(mcs, nr, \texttt{BLER$_t$})
\STATE dci = createDCI(mcs, nr, harq-ack, ...)
\STATE send(ue,dci, NPDCCH)
\STATE // waiting response ... 
\STATE data = receiveFrom(ue,NPUSCH)
\STATE harq-ack = check(data)
\ENDWHILE
\end{algorithmic}
}
\hspace{1pt}
\caption{\texttt{BS} procedure for uplink adaptation.}
\label{alg:BS-THREAD}
\end{algorithm}

Data that is sent to the \texttt{BS} from the \texttt{UE} is partitioned in a sequence of \texttt{TBs}. 
The size in bits of each block is called transfer block size (\texttt{TBS}).
The \texttt{UE} transmits its blocks to the \texttt{BS} through the \texttt{NPUSCH}. 
The success of the block arrival to the \texttt{BS} strongly depends on the signal-to-noise ratio (\texttt{SNR}) of the channel.
Extreme communication scenarios, with respect to \texttt{LTE}, are characterized by low \texttt{SNR} values (less than \SI{-10}{\decibel}~\cite{Yu2017}).
The quality of the communication channel is assessed by the block error rate (\texttt{BLER}) which is defined as the ratio of the number of erroneous blocks received to the total number of blocks sent. 
The \texttt{BS} can estimate the \texttt{BLER} based on the \texttt{ACKs} and \texttt{NACKs} in a given period (e.g. \SI{300}{\ms}~\cite{Yu2017}).

An erroneous block is a \texttt{TB} whose cyclic redundancy check (\texttt{CRC}) is wrong~\cite{ETSI2015}.
In general, high \texttt{BLER} implies many block losses, hence it has to be reduced until a tolerable threshold is reached. 
If the \texttt{SNR} is very low, \texttt{BLER} will be high, unless corrective actions are applied. 
For this purpose, \texttt{NB-IoT} system makes use of two mechanisms: changes to \texttt{MCS} and \texttt{NR}.

The \texttt{MCS} is a number that determines the modulation, which can be \texttt{QPSK} or \texttt{BPSK}~\cite{ETSI2017}.
Furthermore, for a particular \texttt{TBS}, the \texttt{MCS} also defines the coding scheme. 
This scheme is represented as the pipeline: \texttt{CRC}, turbo-coding, and rate-matching~\cite[\S6.3.2]{TS136212}, which transforms the \texttt{TB} at each stage.
In general, a low \texttt{MCS} implies a greater level of redundancy, and therefore, lower \texttt{BLER}. 

\texttt{NR} is the number of transmission repetitions.
High \texttt{NR} also helps to reduce \texttt{BLER} and enhance coverage. 
In the uplink process, \texttt{NR} has the following power-of-two sequence: [1,2,4,...,128].
Another important characteristic of this \texttt{NB-IoT} mechanism is that \texttt{BLER}'s dependency on (\texttt{MCS},\texttt{NR}) is associated with the \texttt{TBS} of each \texttt{TB}.
Summarizing, the relation between the latter parameters responds to the function \texttt{\emph{BLER(TBS,SNR,MCS,NR)}}.

The use of these two mechanisms has an intrinsic cost: the number of required \texttt{RUs}.
In \texttt{NB-IoT}, one \texttt{RU} is the minimum schedulable unit in \texttt{NPUSCH} transmissions.
For \SI{15}{\kilo\hertz} it consists of two slots (\SI{1}{\ms}) for $12$ sub-carriers, four slots (\SI{2}{\ms}) for six sub-carriers, eight slots (\SI{4}{\ms}) for three sub-carriers, $16$ slots (\SI{8}{\ms}) for a single sub-carrier; and for a single sub-carrier of \SI{3.75}{\kilo\hertz}, $16$ slots (\SI{32}{\ms})~\cite{Yu2017}. 
Depending on the \texttt{MCS} assigned by the \texttt{BS}, each \texttt{TB} is coded, in turn, in a sequence of \texttt{RUs}.
The relation between \texttt{TBS}, \texttt{MCS}, and \texttt{RUs}, disregarding repetitions, is established in Table~\ref{table:uplink} and can be formalized by the following function: \texttt{\emph{RUs$_{\text{no-rep}}$(TBS, MCS)}}.

\begin{table}[h]
\centering
\scalebox{0.92}{
\begin{tabular}{|c|c|c|c|c|c|c|c|c|}
\hline
\diagbox[width=\dimexpr \textwidth/8+2\tabcolsep\relax, height=0.7cm]{  MCS $\equiv$ $I_{TBS}$ }{$RUs_{no-rep}$}
                    & 1 & 2 & 3 & 4 & 5 & 6 & 8 & 10  \\ \hline \hline
0 $\equiv$ 0 & 16 & 32 & 56 & 88 & 120 & 152 & 208 & 256 \\ \hline
2 $\equiv$ 1 & 24 & 56 & 88 & 144 & 176 & 208 & 256 & 344 \\ \hline
1 $\equiv$ 2 & 32 & 72 & 144 & 176 & 208 & 256 & 328 & 424 \\ \hline
3 $\equiv$ 3 & 40 & 104 & 176 & 208 & 256 & 328 & 440 & 568 \\ \hline
\end{tabular}
}
\caption{Payload (\texttt{TBS}) for different combinations of coding scheme for \texttt{NPUSCH}, in single sub-carrier uplink~\cite{ETSI2017}.}
\label{table:uplink}
\end{table}

In extreme coverage scenarios, where \texttt{SNR} is very low, achieving small values of \texttt{BLER} (modifying \texttt{MCS} and \texttt{NR}) has an important \texttt{RU} cost. 
To obtain the total number \texttt{RUs}, the number of repetitions has to be considered by multiplying it by \texttt{\emph{RUs$_{\text{no-rep}}$(TBS, MCS)}}.

\section{Link adaptation strategies}\label{sec:uplink-scheduling}

In this section, we analyze the resource usage and the impact of convergence for any link adaptation strategy.
Then we introduce our proposed \texttt{LUT}-based strategy jointly with a set of alternative ones used for evaluation.

\subsection{Resource usage in \texttt{(MCS,NR)} scheduling} \label{subsec:ITBS-NR-0}

The parameters \texttt{MCS} and \texttt{NR} contain information about the modulation and coding scheme, the number of used \texttt{RU}, and the number of repetitions. 
Their values affect the resource usage of the \texttt{NPUSCH}, which is characterized by \texttt{RU} consumption and is directly associated with the number of devices capable of transferring information to the \texttt{BS}.
As uplink \texttt{HARQ} retransmissions are stopped when a maximum threshold is reached~\cite{Huawei2017}, this analysis considers two main scenarios: unacknowledged and acknowledged services, where retransmissions are disabled and enabled, respectively.

In an unacknowledged service, if a block is lost due to transmission problems (for example due to high noise in the channel), no attempt is made to detect the loss or to recover it. 
This class of service is appropriate when \texttt{BLER} is very low, so recovery is left to higher layers. 
It is also suitable for time-critical applications, such as real-time traffic, where having late data is worse than having bad data~\cite{Tanenbaum2010}.  
In this kind of service, decreasing \texttt{RU} cost can be achieved through the selection of an \texttt{(MCS,NR)} tuple which increases the incoming \texttt{BLER} and the block losses. 
Despite that it degrades connection quality (\texttt{BLER} is not the lowest), an \texttt{RU} consumption decrease is associated with a reduction in \texttt{BS} waiting time, which in turn, causes an improvement over \texttt{NPUSCH} resource usage. 
Therefore, this technique can help to provide service to a larger number of devices.

In an acknowledged service, each block sent by the \texttt{UE} is individually acknowledged by the \texttt{BS}. 
If the block has not arrived within a specified time interval, it can be sent again. 
This kind of service is useful for noisy channels, such as cellular systems. 
The downside of this strategy is that it can be inefficient, but on (inherently unreliable) wireless channels it is well worth the cost~\cite{Tanenbaum2010}.

In the latter case, in order to estimate resource usage, \texttt{RU} cost needs to be calculated. 
If current \texttt{BLER} is equal to zero, the probability of \texttt{TB} successful arrival is $1$, which means that no retransmission is required. 
Therefore the \texttt{RU} cost for this transmission is basically \texttt{RUs}. 
On the other hand, if \texttt{BLER} is greater than zero, the probability of the \texttt{TB} arrival is more complex due to retransmissions. 
In particular, we analyze the limiting case in which the maximum retransmission threshold (\texttt{N}) is as high as needed, i.e \texttt{TB} will be transmitted repeatedly until its arrival is successful. 
In every $i$-th transmission, the probability of successful arrival $(1-\texttt{BLER}_i)$ increases due to \texttt{HARQ} mechanism. 
Information about the latest transmissions is used at the \texttt{BS} to decrease \texttt{BLER}, \textit{ergo}: $\texttt{BLER}_i > \texttt{BLER}_{i+1}$.
Therefore, in a constant \texttt{SNR} channel, the expected \texttt{RU} cost is:
\begin{equation}
\overline{\texttt{RUs}} = \texttt{RUs} \sum_{i=0}^{N-1} \left\{ ( i + 1 )\left(\prod_{j=0}^{i-1}\texttt{BLER}_j\right)(1-\texttt{BLER}_i) \right\}
\label{Expected-RU}
\end{equation}
Here, the technique consists in determining the tuple \texttt{(MCS,NR)} which approximates \texttt{BLER} to $\texttt{BLER}_{t}$, which in turn minimizes $\overline{\texttt{RU}}$ cost. 
If this cost is decreased, resource usage will be decreased as well.

\subsection{Convergence in \texttt{(MCS,NR)} scheduling}\label{subsec:ITBS-NR-1}

The strategies that will be presented are single sub-carrier scheduling schemes which dynamically determines \texttt{MCS} and \texttt{NR} for reducing \texttt{BLER} until \texttt{BLER$_t$} is reached.
The first approach to this problem was proposed in Yu et al.~\cite{Yu2017}, wherein each rescheduling \texttt{BLER} is estimated and then a tuple \texttt{(MCS,NR)} is determined with the aim of approximating $\texttt{BLER}_{t}$ (which they fix to 10\%).
Since \texttt{BLER} also depends on current \texttt{SNR} and \texttt{TBS}, there are many tuples \texttt{(MCS,NR)} that could achieve the target threshold: $ candidates = \{ \texttt{(MCS,NR)} \setminus \texttt{BLER(TBS,SNR,MCS,NR)} = \texttt{BLER}_{t} \}$.
Only in the scenario in which the algorithm converges to \texttt{BLER$_t$} the number of block losses is the expected one. 
If the number of options (i.e. \texttt{candidates} size) is large, convergence time will be probably high increasing the number of blocks losses. 
Therefore, convergence time is an important parameter to be minimized.

Absolute convergence to $\texttt{BLER}_{t}$ is unlikely, because (\texttt{MCS},\texttt{NR}) are discrete numbers, which are associated with discrete changes in \texttt{BLER}. 
The probable scenario is an oscillatory one, in which \texttt{BLER} cannot reach $\texttt{BLER}_{t}$ but oscillates around it. 
An important issue occurs when oscillations are big enough to set block losses far above or below the pretended tolerable value. 
In particular, the oscillatory scenario where block losses are high is worsened when retransmissions are enabled because all lost blocks need to be retransmitted.

\subsection{Alternative Strategies}\label{subsec:competitor-algorithms}

Before proposing our \texttt{LUT}-based strategy, in order to help to analyze its performance, a set of alternative strategies is presented. 
They use the previously detailed technique based on the convergence to \texttt{BLER$_t$}, although they do not try to minimize convergence time.
Table~\ref{table:uplink} presents an extract of the available payload for some combinations of coding scheme and number of resource units used.
As can be seen, the variation of \texttt{MCS} is not directly related with an increased level of redundancy. 
For this reason and following the standard practice in this area, the transport block size index \texttt{$I_{TBS}$} is used during the rest of the work.

\begin{itemize}[leftmargin=*]

\item \texttt{$I_{TBS}$-NR}: in this first strategy (Alg.~\ref{alg:ITBS-NR}), each time re-scheduling is executed, \texttt{BLER} is estimated. 
\texttt{BLER} surpassing $\texttt{BLER}_{t}$ suggests that channel quality is poor and \texttt{$I_{TBS}$} is decreased. 
If \texttt{$I_{TBS}$} cannot be decreased any more, \texttt{NR} is increased. 
On the other hand, if the estimated value falls bellow $\texttt{BLER}_{t}$, it means that too many \texttt{RUs} are being used and opposite corrective actions need to be taken.

\begin{algorithm}
{\fontsize{8.5}{8.5}\selectfont
\begin{algorithmic}[1]
\REQUIRE $TBS, BLER_{t}$
\ENSURE $(I_{TBS},NR) \ \in \ candidates, \ when \ iterations \rightarrow \ \infty$
\REPEAT 
  \STATE $Estimate \  BLER$ 
  \IF{$BLER > BLER_{t}$}
    \IF{$I_{TBS} > I_{TBS}^{min}$}
      \STATE $I_{TBS} \leftarrow I_{TBS} - 1$
    \ELSIF{$NR < NR^{max}$}
      \STATE $NR \leftarrow NR * 2$
    \ELSIF{}
      \STATE Bad\ channel\  quality.
      \STATE Target BLER can't be achieved.
    \ENDIF

  \ELSIF{$BLER < BLER_{t}$}
    \IF{$I_{TBS} < I_{TBS}^{max}$}
      \STATE $I_{TBS} \leftarrow I_{TBS} + 1$
    \ELSIF{$NR > NR^{min}$}
      \STATE $NR \leftarrow NR / 2$
    \ELSIF{}
      \STATE Good channel quality.
      \STATE RU consumption can't be decreased.
    \ENDIF
  \ELSIF{}
    \STATE BLER is in range.
  \ENDIF
\UNTIL{$BLER = BLER_{t}$}
\COMMENT{re-scheduling is not needed}
\end{algorithmic}
}
\hspace{1pt}
\caption{\texttt{$I_{TBS}$-NR}}
\label{alg:ITBS-NR}
\end{algorithm}

\end{itemize}

Analogously, we can define other strategies to be used as a base of comparison:

\begin{itemize}[leftmargin=*]
\item \texttt{NR-$I_{TBS}$}:  this is the inverse case of \texttt{$I_{TBS}$-NR} approach; i.e., when estimated \texttt{BLER} is poor, \texttt{NR} is increased. 
When the \texttt{NR} limit is achieved, \texttt{$I_{TBS}$} is modified.

\item \texttt{$I_{TBS}$}: only \texttt{$I_{TBS}$} is used. \texttt{NR} is set to a fixed value.

\item \texttt{NR}: similar to the strategy above, here only \texttt{NR} is used. \texttt{$I_{TBS}$} is fixed to the average value (\texttt{$I_{TBS}^{max}$} / 2).

\item \texttt{$I_{TBS}$\&NR}:  in this approach both, \texttt{$I_{TBS}$} and \texttt{NR}, are changed at the same time; e.g., when measured \texttt{BLER} is high, \texttt{$I_{TBS}$} is decreased in one point, and \texttt{NR} is duplicated.
\end{itemize}

\subsection{\texttt{LUT}-based strategy (\texttt{LUTS})}\label{sec:LUTS-algorithm}

\begin{figure}[!htb]
    \centering
    \includegraphics[width=0.49\textwidth]{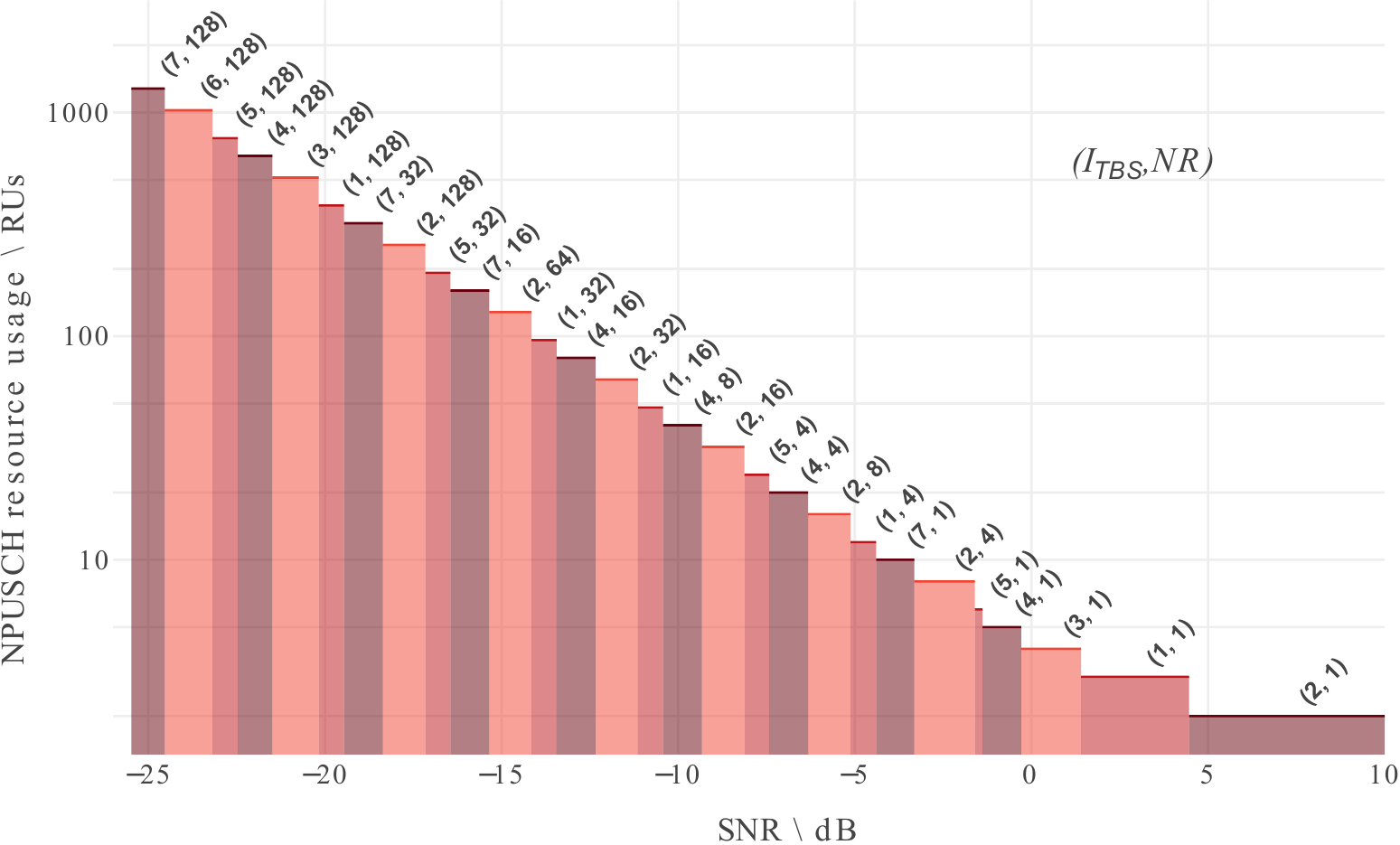}
    \caption{\texttt{RUs} vs \texttt{SNR} for different optimal tuples (\texttt{I$_{TBS}$,NR}).}
    \label{fig:RU_vs_SNR}
\end{figure}

In the traditional approach, for each block transferred from the \texttt{UE}, the \texttt{BS} determines a tuple \texttt{($I_{TBS}$,NR)}.
This process is iterative and converges when a target \texttt{BLER} is achieved. 
During those iterations many transfer blocks could be lost, requiring retransmissions which, in turn, increase the resource usage.
In order to reduce this \texttt{RU} consumption, we propose replacing the described iterative process by memory use, i.e. a lookup table (\texttt{LUT}).
Our \texttt{LUT}-based strategy (\texttt{LUTS}), needs three input parameters: the \texttt{TBS}, the \texttt{SNR}, and the $\texttt{BLER}_{t}$.
The first parameter, the \texttt{TBS}, is known by the \texttt{BS} and \texttt{UE}, the \texttt{BS} impose this magnitude at the beginning of each transmission.
The second parameter is the \texttt{SNR}, unlike the former strategies which use a \texttt{BLER} estimation. 
An \texttt{SNR} estimation is taken after the equalization and compensation stages (a comparison of \texttt{SNR} estimation techniques is shown in Pauluzzi et al.~\cite{Pauluzzi2000}).
The estimated magnitude can absorb some level of uncertainty. 
We exemplify this in Fig.~\ref{fig:RU_vs_SNR}, for a \texttt{TBS} of \SI{256}{\bit}, we show how the \texttt{SNR} varies with respect to the consumed \texttt{RUs}.  
As can be seen, the effective \texttt{SNR} estimation can absorb uncertainties up to \SI{0.5}{\decibel} or even \SI{1}{\decibel} in some ranges without modifying \texttt{RU} consumption.

The last parameter is the $\texttt{BLER}_{t}$, thus the target (or tolerable) block error rate.
For any given pair \texttt{TBS} and \texttt{SNR}, there are just a few possible $\texttt{BLER}_{t}$ values available.
In the following \texttt{LUT} extract (see Table \ref{table:LUT_extract}), it can be seen that \texttt{BLER} jumps from near 0 to $0.7$, and then reaches almost $1$. 
Each available $\texttt{BLER}_{t}$ or $\texttt{BLER}$ range represents a \texttt{QoS}.

\begin{table}[h]
\centering
\begin{tabular}{|c|c|c|c|c|c|c|}
\hline
\texttt{TBS} \textbackslash bits & \texttt{SNR} \textbackslash dB & \texttt{BLER$_t$} & \texttt{QoS} &\texttt{$I_{TBS}$} & \texttt{NR}  & \texttt{RUs}  \\ 
\hline  \hline 
256          & -24          & \cellcolor[HTML]{FFF5F0}{\color[HTML]{000000} 0.00006} & \cellcolor[HTML]{FFF5F0}good   & 0             & 128         & 1280         \\ \hline
256          & -24          & \cellcolor[HTML]{FFF5F0}{\color[HTML]{000000} 0.00491} & \cellcolor[HTML]{FFF5F0}good   & 1             & 128         & 1024         \\ \hline
256          & -24          & \cellcolor[HTML]{FDCAB5}{\color[HTML]{000000} 0.70477} & \cellcolor[HTML]{FDCAB5}poor & 2             & 128         & 768          \\ \hline
256          & -24          & \cellcolor[HTML]{F1AC90}{\color[HTML]{000000} 0.98001} & \cellcolor[HTML]{F1AC90}bad    & 0             & 64          & 640          \\ \hline
256          & -24          & \cellcolor[HTML]{F1AC90}{\color[HTML]{000000} 1}       & \cellcolor[HTML]{F1AC90}bad    & 0             & 1           & 10           \\ \hline
\end{tabular}
\caption{\texttt{LUT} extract sample. Some of the \texttt{BLER} values can be collapse into a single QoS.}
\label{table:LUT_extract}
\end{table}

Based on this information, \texttt{LUTS} can use the lookup table to retrieve the \texttt{$I_{TBS}$} and \texttt{NR}. 
If the row containing the \texttt{TBS}, estimated \texttt{SNR}, and $\texttt{BLER}_{t}$ exists then its associated tuple: ($\texttt{$I_{TBS}$}^{opt}$,$\texttt{NR}^{opt}$) is the optimal row, providing the minimum \texttt{RU} consumption under these conditions.
Nevertheless, in some cases, the row could not be found and an approximation policy has to be implemented.
This policy consists of retrieving the row with \texttt{BLER} and \texttt{SNR} values less or equal to $\texttt{BLER}_{t}$ and the estimated \texttt{SNR}, respectively, with the addition of minimum \texttt{RU} cost.
\texttt{LUTS} is reduced to the following straightforward pseudo-code presented in Alg.~\ref{LUTS}.

\begin{algorithm}
\caption{The LUT-based strategy selects the optimal codification and number of repetitions based on the estimated \texttt{SNR} and the selected $\texttt{BLER}_{t}$ or \texttt{QoS}.}
\label{LUTS}
{\fontsize{9}{9}\selectfont
\begin{algorithmic}[1]
\REQUIRE $TBS, BLER_{t}, \texttt{LUT}$
\STATE $Estimate \  SNR$
\STATE $SNR^{new}, BLER_{t}^{new} \leftarrow \texttt{LUT}.getClosestMinRU(TBS,SNR,BLER_{t}) $
\STATE $I_{TBS}, NR \leftarrow \newline
\texttt{LUT}.getTuple(TBS, {SNR}^{new}, BLER_{t}^{new})$
\end{algorithmic}
}
\hspace{1pt}
\end{algorithm}

\subsection{\texttt{LUT} initialization}\label{sec:LUTS-LUT-algorithm}

A \texttt{LUT} initialization method is proposed. 
An exploratory algorithm (Alg.~\ref{alg:LUTS-LUT-CALC}) begins with the \texttt{LUT} uninitialized. 
Every time a \texttt{UE} is connected, the algorithm estimates the \texttt{SNR}. 
A tuple \texttt{$(I_{TBS}$,NR)} is obtained using one of the strategies presented in the previous section (e.g. \texttt{$I_{TBS}$-NR}), which guarantees a \texttt{BLER} closer to $\texttt{BLER}_{t}$. 
If the row with key \texttt{(TBS,SNR,\texttt{BLER}$_{t}$)} does not exist, it is added to the \texttt{LUT}.
If it exists, \texttt{RU} consumption is compared between new and old \texttt{$(I_{TBS}$,NR)}. 
The \texttt{LUT} is updated with the tuple with minimum resources (which is also the tuple that is used for the transmission). 
Ideally, this process is repeated until each row converges to minimum \texttt{RUs}, i.e., \texttt{LUT} is completed. 
The \texttt{BS} needs a large number of connections from different type to populate the \texttt{LUT}. 
Furthermore, as it was mentioned in the previous section, few \texttt{QoS} are presented in the \texttt{LUT}, therefore a limited number of \texttt{BLER} ranges are reasonable to be used (e.g. `good').
On the other hand, a more pragmatic approach could use physical layer simulations, like the ones used in our simulator, to create a pre-calculated \texttt{LUT}. 
In this case, the \texttt{BS} can use this table and update it during the interaction with the devices.
Moreover, concerning \texttt{LUT} size, it ranges from some kilobytes (e.g. \SI{200}{KB}) to some megabytes (e.g. \SI{3.4}{MB}), depending on the implementation. 
Finally, even though the initialization is time-consuming, it is also a one-time process.

\begin{algorithm}
\caption{\texttt{LUT} initialization.}
\label{alg:LUTS-LUT-CALC}
{\fontsize{8.5}{8.5}\selectfont
\begin{algorithmic}[1]
\REQUIRE $TBS, BLER_{t}$
\ENSURE $\texttt{LUT} \ is \ complete, \ when \ iterations \rightarrow \ \infty$
\REPEAT 
\STATE $Estimate \  SNR$
\STATE $I_{TBS}^{new},NR^{new} \leftarrow ITBS\text{-}NR(BLER_{t}) \ $
\IF{ $\texttt{LUT}.rowExists(TBS,SNR,BLER_{t})$} 
\STATE $I_{TBS}^{old},NR^{old}$
\STATE $\ \ \ \ \leftarrow \texttt{LUT}.getTuple(TBS,BLER_{t},SNR)$
\IF{$RUs(I_{TBS}^{new},NR^{new})<RUs(ITBS^{old},NR^{old})$}
\STATE $\texttt{LUT}.setRow(TBS,SNR,BLER_{t},I_{TBS}^{new},NR^{new})$
\ENDIF
\ELSE
\STATE $\texttt{LUT}.setRow(TBS,SNR,BLER_{t},I_{TBS}^{new},NR^{new})$
\ENDIF
\UNTIL{\texttt{LUT} is complete}
\end{algorithmic}
}
\hspace{1pt}
\end{algorithm}

\section{Simulation} \label{sec:simulation}

In order to test the proposed strategies, we introduce an open source \texttt{NPUSCH} uplink simulator. 
It models the uplink iterative sub-process, where the \texttt{BS} determines the \texttt{($I_{TBS}$, NR)} tuple and sends this information to the \texttt{UE}.
The software is implemented in \texttt{Python} and \texttt{LUT} values are obtained from a simulation based on the \texttt{NB-IoT Uplink Waveform Generation} from the \texttt{Matlab Toolkit}.
In this simulation, for each (\texttt{$I_{TBS}$}, \texttt{$I_{RU}$}, \texttt{$I_{REP}$}) 3-tuple, \texttt{BLER} curves were traced for a single sub-carrier mode \texttt{NPUSCH} over a simulated \texttt{AWGN} channel. 
The \texttt{TX-RX} chain is based on the standard specifications (\cite[\S10.1.3]{TS136211} \& \cite[\S6.3.2]{TS136212}). 
Project source code can be found in~\cite{NBIoTSim2018}. 
Preliminary results generated using a previous version of this simulator were presented at~\cite{Lujan2018}.

All proposed scheduling algorithms in this work were implemented in our simulator. 
In all the experiments, it is assumed that the \texttt{LUT} initialization stage had been completed before performing the experiment.

\texttt{LUTS} algorithm is designed to reduce \texttt{BLER} convergence interval. 
The larger this interval is, compared to the number of \texttt{TB} that must be sent, the greater the gain in resource usage. 
In this context, the experiments consisted of $500$ realizations of a \texttt{UE} transmitting to the \texttt{BS} blocks of \SI{256}{\bit}, i.e, the magnitude of a possible alarm message. 
Extreme coverage scenarios were characterized by \texttt{SNRs} of \SIlist{-24;-20;-16}{\decibel}. 
Analyzed \texttt{BLER$_t$} corresponds to a `good' \texttt{QoS} (in section \ref{sec:results-and-discussion} a discussion about this value is presented).

\section{Results and Discussion}\label{sec:results-and-discussion}

\subsection{Unacknowledged Service}

\begin{figure}[!htb]
    \centering
    \includegraphics[width=0.49\textwidth]{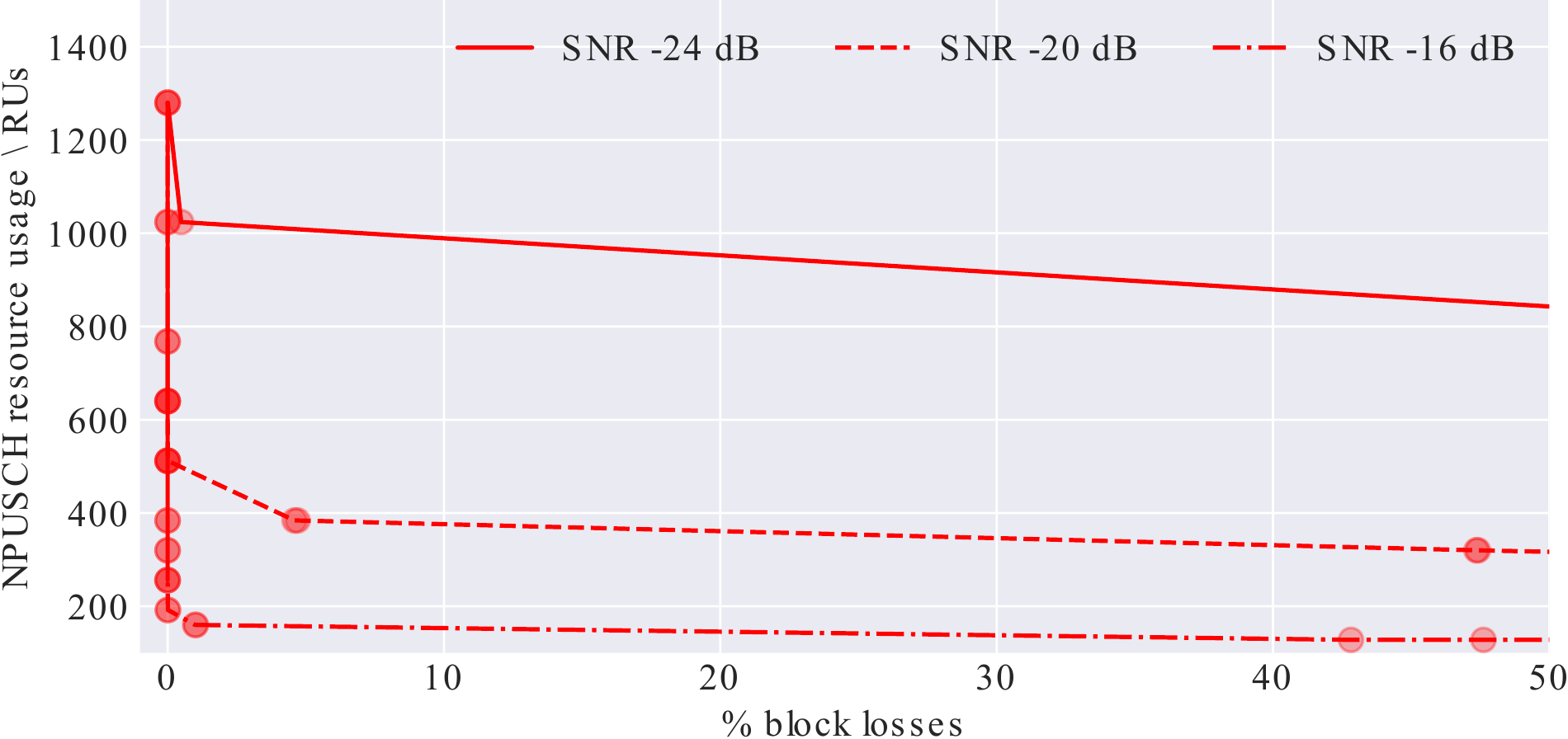}
    \caption{\texttt{NPUSCH} resource usage {\it vs} percentage of block losses for different extreme coverage scenarios, in unacknowledged mode.}
    \label{fig:nru_bler_nonret}
\end{figure}

Based on the outcome provided by our uplink simulator, we obtain a table with the structure previously showed in Table~\ref{table:LUT_extract} (which can be found in the simulator repository~\cite{NBIoTSim2018}).
This information is used to calculate the relation between \texttt{NPUSCH} resource usage (measured in \texttt{RUs}) and the percentage of block losses (\texttt{BLER}*$100$). 
This relation is presented in Fig.~\ref{fig:nru_bler_nonret}, considering different extreme coverage scenarios, represented by low \texttt{SNRs}. 
This figure shows that the \texttt{RU} cost significantly diminishes between $0\%$ and $5\%$ of block losses, and then remains without strong variations. 
We found that a \texttt{BLER$_t$} of $0.05$ is an adequate trade-off between losses and resource usage, thus a `good' \texttt{QoS}.
In particular, when calculating the average of the differences (\texttt{\#RU(BLER=0) - \#RU(BLER=0.05)}) between the \texttt{SNRs}, \texttt{NPUSCH} resource usage is reduced to $63\%$.

Savings mentioned above are only possible when the algorithm that selects the tuple \texttt{($I_{TBS}$,NR)} converges instantaneously.
In a more realistic scenario, this convergence takes some iterations as can be seen next. 

\begin{figure*}
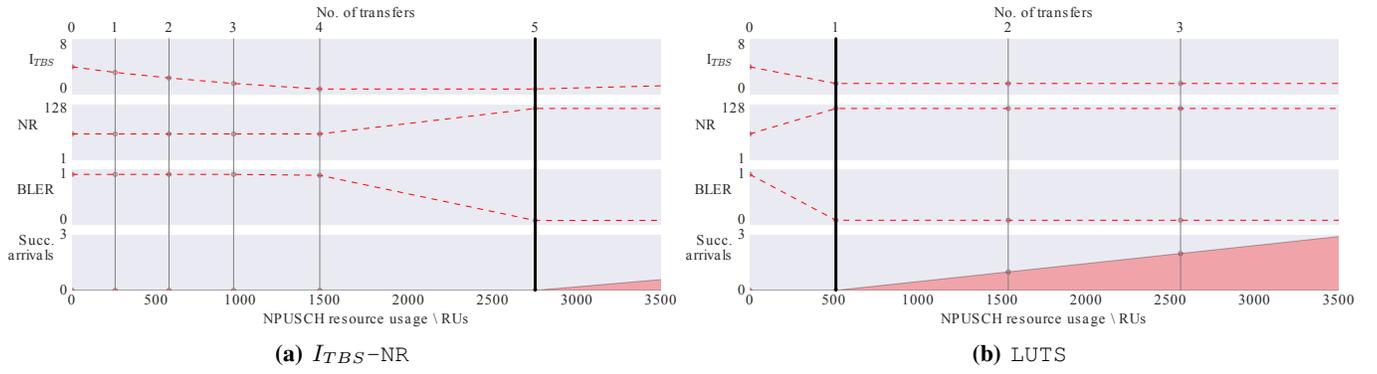

    \centering
    \begin{subfigure}{0.49\textwidth}
      \includegraphics[width=\textwidth]{{{ITBS-NR_targetBLER_5_SNR_-24_ret_False}}}
      \caption{\texttt{$I_{TBS}$-NR}} 
      \label{fig:ITBS-NR-NORET}
    \end{subfigure}
    \begin{subfigure}{0.49\textwidth}
      \includegraphics[width=\textwidth]{{{LUTS_targetBLER_5_SNR_-24_ret_False}}}
      \caption{\texttt{LUTS}} \label{fig:LUTS-NORET}
    \end{subfigure}
    \caption{ \texttt{$I_{TBS}$-NR} and \texttt{LUTS} algorithms evolution in extreme coverage condition of \SI{-24}{\decibel} \texttt{SNR}. 
    The x-axes show the number of transferred blocks (top) and the accumulated \texttt{RUs} (bottom). Both axes have a correspondence with time.
    The y-axes show main algorithm parameters: \texttt{$I_{TBS}$}, \texttt{NR}, \texttt{BLER} and the No. of successful arrivals.} 
    \label{fig:algorithms-evolution-noret}\label{fig:ALGO-EVOL-NORET}
\end{figure*}

Fig.~\ref{fig:ALGO-EVOL-NORET} presents the evolution of two previously described algorithms: \texttt{$I_{TBS}$-NR} and \texttt{LUTS}. 
The top x-axis shows the number of transferred blocks, the bottom x-axis shows the \texttt{NPUSCH} resource usage measured in terms of the accumulated \texttt{RUs}. 
These axes also have a correspondence with time, which is not depicted in the figures because it would imply the addition of downlink channels information, not being addressed in this work.
In both figures, each circle tags the time of a new transmission from the \texttt{UE} to the \texttt{BS}. 
Each figure is composed of four gray subfigures.
The first two subfigures show how \texttt{$I_{TBS}$} and \texttt{NR} vary. 
The third subfigure shows how \texttt{BLER}, measured at the \texttt{BS}, varies at each new transmission. 
Finally, the fourth subfigure shows the number of successful transmissions at each moment.
In the experiment shown in Fig.~\ref{fig:ALGO-EVOL-NORET}, $20$~\texttt{TBs} with \texttt{TBS} of \SI{256}{\bit} are sent from the \texttt{UE} to the \texttt{BS}, under a constant extreme coverage condition of \SI{-24}{\decibel} \texttt{SNR}. 
It can be observed that while \texttt{$I_{TBS}$-NR} took five transmissions to converge to the optimal tuple \texttt{($I_{TBS}=0$,NR$=128$)}, \texttt{LUTS} took only one iteration due to the use of pre-calculated values.
Furthermore, in this particular experiment, when approximately $3500$~\texttt{RUs} were used, \texttt{LUTS} achieved three successful transmissions, while \texttt{$I_{TBS}$-NR} none.
These savings are better analyzed next.

\begin{figure*}
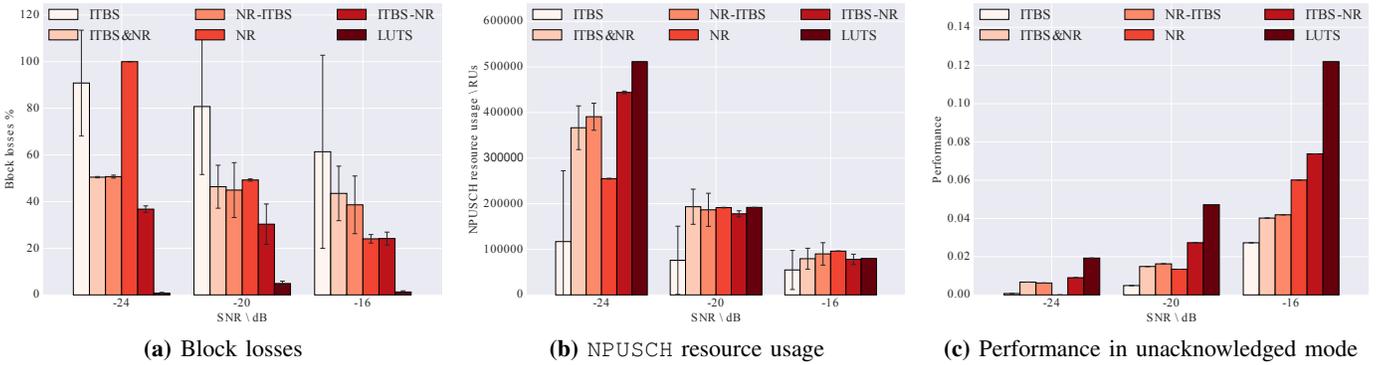

    \begin{subfigure}{0.32\textwidth}
        \includegraphics[width=\textwidth]{{{losses_snr_sched_bler_5_ret_False}}}
        \caption{Block losses}
        \label{fig:BL-BLER03}
    \end{subfigure}
    \hfill
    \begin{subfigure}{0.32\textwidth}
        \includegraphics[width=\textwidth]{{{nru_snr_sched_bler_5_ret_False}}}
        \caption{\texttt{NPUSCH} resource usage}
        \label{fig:LAT-BLER03}
    \end{subfigure}
    \hfill
    \begin{subfigure}{0.32\textwidth}
        \includegraphics[width=\textwidth]{{{performance_snr_sched_bler_5_ret_False}}}
        \caption{Performance in unacknowledged mode}
        \label{fig:P-BLER03}
    \end{subfigure}
    \caption{\texttt{UE} sends $500$ blocks of \SI{256}{\bit} each to the \texttt{BS} using `good' \texttt{QoS} under different \texttt{SNRs}. }
    \label{fig:noret}
\end{figure*}

In Fig.~\ref{fig:noret} the previous experiment is extended with all the algorithms presented in sub-section \ref{subsec:competitor-algorithms}, and with the different extreme coverage \texttt{SNR} values presented in section \ref{sec:simulation}. 
Three parameters were analyzed: block losses, resource usage, and performance.  
Subfigure~\ref{fig:BL-BLER03} shows that \texttt{LUTS} is the algorithm that minimizes block losses in all cases.
It only loses less than $10\%$ of the blocks, while the other algorithms lose $35\%$ or more. 
Subfigure~\ref{fig:LAT-BLER03} shows the resource usage of \texttt{NPUSCH}, \texttt{LUTS} is one of the algorithms that consume more \texttt{RUs}. 
Consequently, a performance parameter is needed to compare the algorithms considering their impact on block losses and resource usage simultaneously. 
To address this, we introduce \texttt{P} which represents the performance:
\begin{equation}
\texttt{P} = \texttt{Efficiency} \cdot (1 - \texttt{BLER}) = \frac{ ( 1 - \texttt{BLER} )^2 }{ \texttt{\#RU}}\label{performance}
\end{equation}
The efficiency is defined by the quotient between the proportion of successful block arrivals by the transmission cost (\texttt{$\#$RUs}). 
The efficiency could present high values even if only a small number of blocks can be successfully transmitted. 
To consider the number of transmissions and successfully received blocks simultaneously, the efficiency should be multiplied by ($1-$\texttt{BLER}), obtaining the performance (\texttt{P}).
Figure~\ref{fig:P-BLER03} introduces \texttt{P} obtained with all the algorithms. 
\texttt{LUTS} presents the highest performance in all coverage scenarios.

\subsection{Acknowledged Service}

\begin{figure}[!htb]
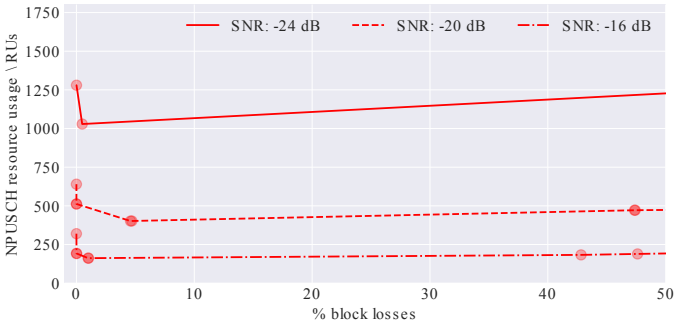

    \includegraphics[width=0.49\textwidth]{{{nru_bler_ret}}}
    \caption{\texttt{NPUSCH} resource usage {\it vs} percentage of block losses for different extreme coverage scenarios in acknowledged mode.}
    \label{fig:nru_bler_ret}
\end{figure}

Fig.~\ref{fig:nru_bler_ret} is analogous to Fig.~\ref{fig:nru_bler_nonret}, with the addition of retransmissions. 
In this case, Eq.~\ref{Expected-RU} is necessary to calculate \texttt{NPUSCH} resource usage. 
For all the analyzed block losses percentages, the number of retransmissions remained near three.
As before, $5\%$ of block losses provides a reasonable trade-off between \texttt{RU} usage and block losses, which represents a `good' \texttt{QoS}.
This time \texttt{NPUSCH} resource usage was reduced to $56.7\%$, calculated as the average of the differences (\texttt{\#RU(BLER=0) - \#RU(BLER=0.05)}) between the \texttt{SNRs}. 
Analogously to Fig.~\ref{fig:nru_bler_nonret}, this result is only valid when tuple \texttt{($I_{TBS}$,NR)} convergence is instantaneous.
The algorithms analyzed in the figures below propose more realistic scenarios, where convergence time is not null.

\begin{figure*}
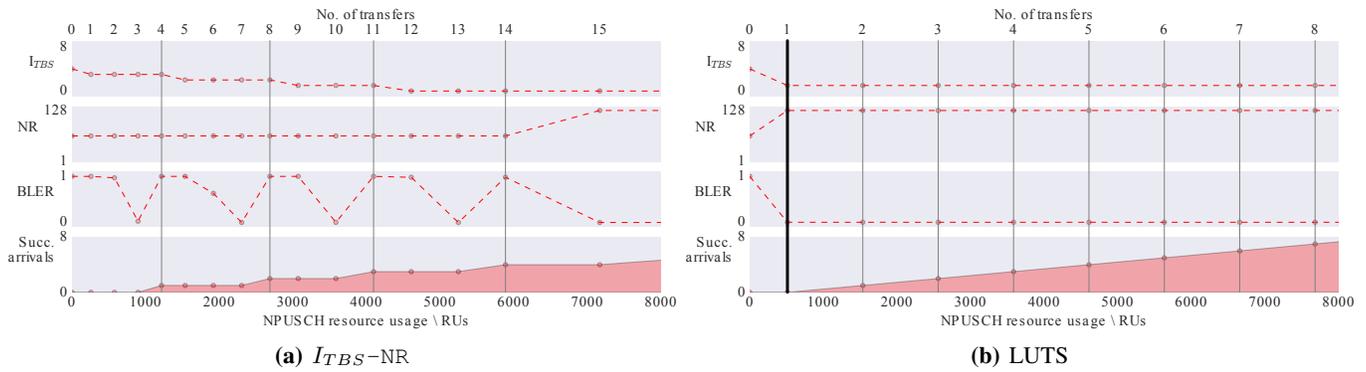

    \centering
    \begin{subfigure}{0.49\textwidth}
      \includegraphics[width=\textwidth]{{{ITBS-NR_targetBLER_5_SNR_-24_ret_True}}}
      \caption{\texttt{$I_{TBS}$-NR}}
      \label{fig:ITBS-NR}
    \end{subfigure}
    \begin{subfigure}{0.49\textwidth}
      \includegraphics[width=\textwidth]{{{LUTS_targetBLER_5_SNR_-24_ret_True}}}
      \caption{LUTS}
      \label{fig:LUTS}
    \end{subfigure}
    \caption{\texttt{$I_{TBS}$-NR} (a) and \texttt{LUTS} (b) evolution in acknowledged mode. The x-axes show the number of transferred blocks (top) and the accumulated \texttt{RUs} (bottom). Both axes have a correspondence with time.
    The y-axes show main algorithm parameters: \texttt{$I_{TBS}$}, \texttt{NR}, \texttt{BLER} and the No. of successful arrivals. }
    \label{fig:ALGO-EVOL-RET}
\end{figure*}

\begin{figure}[!htb]
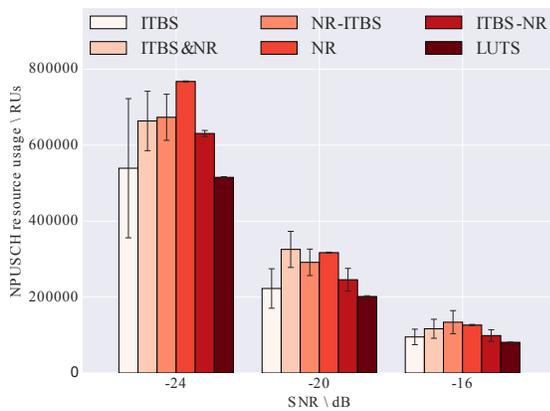

	\centering
    \includegraphics[width=0.4\textwidth]{{{nru_snr_sched_bler_5_ret_True}}}
    \caption{\texttt{NPUSCH} resource usage (average and standard deviation) in acknowledged mode, for different \texttt{SNRs} and \texttt{QoS}: `good'. \texttt{UE} sends $500$ blocks of \SI{256}{\bit} each to the \texttt{BS}.}
    \label{fig:ret}
\end{figure}

Fig.~\ref{fig:ALGO-EVOL-RET} presents the evolution of the algorithms \texttt{$I_{TBS}$-NR} and \texttt{LUTS} when retransmissions are available. 
This figure is similar to Fig.~\ref{fig:ALGO-EVOL-NORET}. 
Here, \texttt{$I_{TBS}$-NR} needs $15$ transmissions to converge to optimal (\texttt{$I_{TBS}$}, \texttt{NR}) tuple while \texttt{LUTS} needs only one.
Additionally, in the third subfigure of \texttt{$I_{TBS}$-NR}, during convergence stage, \texttt{BLER} oscillations emerged. 
This behavior was described in section~\ref{subsec:ITBS-NR-1}. 
Furthermore, in the fourth subfigure, when near $8000$~\texttt{RUs} were consumed \texttt{LUTS} successfully transmit seven blocks, while \texttt{$I_{TBS}$-NR} only four.
\texttt{LUTS} transmission efficiency is broadly depicted in Fig.~\ref{fig:ret}.

Finally, Fig.~\ref{fig:ret} shows \texttt{NPUSCH} resource usage in different extreme coverage \texttt{SNR} scenarios and for the different algorithms, over an acknowledged service. 
In this experiments all blocks were transmitted as many times as necessary, i.e., all blocks successfully arrived so it was not necessary to analyze performance (Eq.~\ref{performance}).
While in an unacknowledged service (Fig.~\ref{fig:noret}) \texttt{LUTS} resource usage was one of the highest (but with maximum performance); here, \texttt{LUTS} is the algorithm which minimizes this value in all scenarios.
In particular, in the worst case analyzed (\SI{-24}{\decibel}), when an average between consumption differences \break (e.g. \texttt{\#RU(ITBS) - \#RU(LUTS)}) is calculated, \texttt{LUTS} reduces this resource usage consumption by around $28\%$. 

\section{Conclusion} \label{sec:conclusions}

We have proposed a strategy with the aim of improve massive connectivity in \texttt{NB-IoT} under extreme coverage scenarios.
Our technique is based on reducing radio resource usage of shared channels through link adaptation optimization.
Particularly, we use a lookup table to accelerate the convergence of the main link parameters: modulation and coding scheme, as well as the number of repetitions.
For this purpose, a case-specific open source simulator was developed.
An extensive analysis was performed over the \texttt{NPUSCH} using \texttt{SNR} of \SIlist{-24;-20;-16}{\decibel}.
Results show that few target \texttt{BLER} ranges, thus \texttt{QoS}, are actually available at the \texttt{LUT}.
\texttt{SNR} estimation uncertainties are absorbed up to \SI{1}{\decibel} without modifying consumption.
Simulations of six scheduling strategies, composed by $500$ realizations of the transmission of \SI{256}{bit} blocks (which could be generated by an activated alarm) over mentioned extreme coverage \texttt{SNRs} were performed in acknowledged and unacknowledged modes. 
In the first mode, our strategy minimizes resource usage in all scenarios, reducing \texttt{RU} consumption an average of $28\%$ under the most extreme \texttt{SNR}. 
In the second mode, the \texttt{LUT}-based strategy duplicate performance, calculated based on losses and consumption, with respect to the other alternatives on every \texttt{SNR}.
In closing, we expect our proposed strategy could be relevant for future \texttt{BS} software design and will contribute to the extension of massive terminal access in scenarios of extreme coverage.

\section*{Acknowledgment}

The authors would like to thank Consejo Nacional de Investigaciones Cient\'ificas y T\'ecnicas (CONICET) and University of Buenos Aires (UBA), Argentina.
This work was partially funded by CONICET (PUE 22920160100117CO and PIO13320150100020CO), UBA (UBACyT 20020130200096BA), and ANCyPT (PICT-2015-2761).
We specially thank to the editor and reviewers of this article for their valuable contribution.

\ifCLASSOPTIONcaptionsoff
  \newpage
\fi

\small 
\bibliographystyle{IEEEtran}

\bibliography{journalsAbbr,nbiot}

\end{document}